\documentclass[showpacs,aps,prb,reprint,superscriptaddress,preprintnumbers]{revtex4-2}

\usepackage{amsmath}
\usepackage{amssymb}
\usepackage{graphicx}% Include figure files
\usepackage{dcolumn}
\usepackage[dvipsnames]{xcolor}
\usepackage{amsthm,amssymb}
\usepackage{mathrsfs}

\usepackage{color}
\usepackage{endnotes}
\usepackage{bm}
\usepackage{epsfig}
\usepackage{array}
\usepackage{ulem}
\usepackage{soul}

\newcommand{\CsVSb}{{CsV$_{3}$Sb$_{5}$}}
\newcommand{\CCS}{{CsCr$_{3}$Sb$_{5}$}}
\newcommand{\AVSb}{{AV$_{3}$Sb$_{5}$}} 
\newcommand{\parallelsum}{\mathbin{\!/\mkern-5mu/\!}}

\begin{document}

\title{Anomalous Low-temperature Magnetotransport in Kagome Metal\\ \CCS\ under Pressure}

\author{Zikai~Zhou}
\thanks{These authors contributed equally to this work.}
\affiliation{Department of Physics, The Chinese University of Hong Kong, Shatin, Hong Kong, China}
\author{Wenyan~Wang}
\thanks{These authors contributed equally to this work.}
\affiliation{Department of Physics, The Chinese University of Hong Kong, Shatin, Hong Kong, China}
\affiliation{State Key Laboratory of Quantum Information Technologies and Materials, The Chinese University of Hong Kong, Shatin, Hong Kong, China}
\author{Deng~Hu}
\affiliation{Centre for Quantum Physics, Key Laboratory of Advanced Optoelectronic Quantum Architecture\\ and Measurement (MOE), School of Physics, Beijing Institute of Technology, Beijing 100081, China}
\affiliation{Beijing Key Laboratory of Quantum Matter State Control and Ultra-Precision Measurement Technology, Beijing Institute of Technology, Beijing 100081, China}
\author{Zheyu~Wang}
\affiliation{Department of Physics, The Chinese University of Hong Kong, Shatin, Hong Kong, China}
\affiliation{State Key Laboratory of Quantum Information Technologies and Materials, The Chinese University of Hong Kong, Shatin, Hong Kong, China}
\author{Ying~Kit~Tsui}
\affiliation{Department of Physics, The Chinese University of Hong Kong, Shatin, Hong Kong, China}
\affiliation{Quantum Science Center of Guangdong-Hong Kong-Macao Greater Bay Area, Shenzhen, China}
\author{Tsz~Fung~Poon}
\affiliation{Department of Physics, The Chinese University of Hong Kong, Shatin, Hong Kong, China}
\author{Zhiwei~Wang}
\affiliation{Centre for Quantum Physics, Key Laboratory of Advanced Optoelectronic Quantum Architecture\\ and Measurement (MOE), School of Physics, Beijing Institute of Technology, Beijing 100081, China}
\affiliation{Beijing Key Laboratory of Quantum Matter State Control and Ultra-Precision Measurement Technology, Beijing Institute of Technology, Beijing 100081, China}
\affiliation{International Centre for Quantum Materials, Beijing Institute of Technology, Zhuhai 519000, China}
\author{Swee~K.~Goh}
\email[]{skgoh@cuhk.edu.hk}
\affiliation{Department of Physics, The Chinese University of Hong Kong, Shatin, Hong Kong, China}
\affiliation{State Key Laboratory of Quantum Information Technologies and Materials, The Chinese University of Hong Kong, Shatin, Hong Kong, China}

\date{\today}

\begin{abstract}

{As a unique kagome superconductor displaying clear signatures of strong electronic correlations,  CsCr$_3$Sb$_5$ has drawn much attention. Its rich temperature-pressure phase diagram features intertwined orders including pressure-induced superconductivity and two density-wave-like phases, making it an outstanding platform to explore the complex coexistence and competition of multiple quantum orders. At around 30~K, which we designate as $T_3$, a possible anomaly manifesting as a hump in the resistivity has been observed, yet its nature remains largely unexplored due to limited supporting evidence from other probes. Here, we conducted systematic magnetotransport experiments under hydrostatic pressure to investigate the nature of this anomaly. Our results reveal an abundance of intriguing magnetotransport signatures below $T_3$, including a non-trivial temperature dependence of the Hall coefficient, multi-band characteristics, and pressure-enhanced anomalous-Hall-like effect. These signatures bear resemblance to those observed in many other correlated materials with distinct origins. These findings suggest the possibility of an additional electronic order in CsCr$_3$Sb$_5$, calling for further detailed investigations.}

\end{abstract}

\maketitle

%%%%%%Introduction%%%%%%%%%%%%%%%%%%
Kagome lattices formed by corner-sharing-triangles feature an electronic band structure characterized by van Hove singularities (vHs), Dirac points, and flat bands, establishing them as
outstanding platforms for investigating the interplay between topologically nontrivial states
and electronic correlations~\cite{Ortiz2019,Ortiz2020,Li2021,Jiang2021,Yin2022,Neupert2021}. These unique attributes make them one of the central focuses of contemporary condensed
matter research. For example, \AVSb\ (A = K, Rb, Cs) has been the subject of intense research since its discovery, driven by its rich phase diagram featuring charge density wave (CDW) order, electronic nematicity, and superconductivity (SC)~\cite{Ortiz2019,Ortiz2020,Ortiz2021,Kiesel2012,Kiesel2013,Wang2013,Nie2022,Li2022,Xu2022,Yin2021,Tan2021,Wang2023,Wang2023b,Yang2020,Du2021,Wang2021a,Zheng2022,Kang2022,Asaba2023,Tan2023,zhou2026,wang2024,zhang2024large}. In the CDW state, S-shape field dependence of the Hall effect occurs near the zero field in all three members of \AVSb~\cite{Yang2020,Yu2021b,Wang2023,wang2023d}.  This anomalous-Hall-like effect (AHLE) has been explained in terms of spontaneous time-reversal symmetry breaking~\cite{Jiang2021,Asaba2023,Guo2022}, Berry curvature effects~\cite{Fu2021,Chapai2023}, or extremely high
mobility carriers~\cite{liu2025a}. Given that the superconducting state emerges from the CDW order, the origin of the AHLE becomes an important topic to understand the properties of \AVSb.

Recently, a new kagome metal isostructural to \AVSb\ has been synthesized. When the vanadium kagome nets are replaced by the chromium counterparts, \CCS\ is formed. Intriguingly, the Fermi energy in \CCS\ is located near the flat band~\cite{Wu2025,Xu2025,li2025,wang2025,peng2026,Xie2025}, in contrast to the case of \CsVSb\ where the Fermi energy lies near the vHs. Thus, \CCS\ is expected to feature substantially strong electron correlation. Indeed, an abundance of novel properties was found in the new kagome metal, including pressure-induced superconductivity, frustrated magnetism, large electron effective mass, and non-Fermi liquid behavior~\cite{Liu2024}.   
Moreover, two density-wave-like transitions, characterized by phase transition temperatures $T_1$ and $T_2$, have been identified in \CCS\ through multiple probes, including electronic transport, x-ray diffraction, scanning tunneling microscopy (STM), and nuclear magnetic resonance (NMR) spectroscopy~\cite{Liu2024,li2025a,cheng2026}. According to the constructed temperature-pressure phase diagram, these phases interact strongly with superconductivity. Furthermore, the full suppression of both $T_1$ and $T_2$, accompanied by the non-Fermi-liquid transport, hints at the relevance of quantum criticality~\cite{Liu2024}. In addition to these phases, another feature has been observed in the temperature dependence of the resistivity, manifested as an anomalous hump at $T_3$, which at ambient pressure takes the value of $\approx30$~K~\cite{liu2026,peng2026}. This hump feature is already present in the initial report on \CCS, although it is less pronounced and difficult to resolve~\cite{Liu2024}. Recent datasets have revealed a more prominent hump feature at ambient pressure, but there is no detailed discussion. Furthermore, supporting evidence from other probes beyond electrical resistivity remains lacking, leaving its nature open to further investigation.

In this letter, we systematically study the magnetotransport properties related to $T_3$. At ambient pressure and near-ambient pressure (5~kbar), we observe a distinctive upturn in the temperature dependence of the Hall coefficient ($R_H(T)$) when $T_3$ emerges. Below $T_3$, our results reveal a non-linear Hall response which is absent from the initial report. We further employ hydrostatic pressure to track these transport signatures, and find that the behaviour of $R_H(T)$ correlates with $T_3$. Furthermore, the non-linear Hall response can still be observed below $T_3$, but it becomes even sharper compared with the low-pressure case. Thus, the Hall anomaly becomes more anomalous-Hall-like, similar to that seen in \AVSb~\cite{Yang2020,Yu2021b,Wang2023,wang2023d}. These findings suggest the presence of unconventional charge transport intimately tied to $T_3$, offering a valuable insight into the nature of this \ low-temperature transport anomaly in \CCS.

%%%%%%%%%%%Method%%%%%%%%%%%%%
Single crystals of \CCS\ were synthesized by a flux method using binary CsSb as a flux, and characterized by energy dispersive X-ray (see Supplemental Material~\cite{supp}). Three thin flake samples of \CCS\ with similar thickness were mechanically cleaved from the same batch of bulk crystals. Sample K1 was measured at both 5~kbar and 19~kbar, while sample K2 was measured only at 5~kbar and K3 at ambient pressure. Magnetotransport data up to 14~T were collected using a standard Hall bar configuration with six electrodes in a Quantum Design physical property measurement system (PPMS), with temperatures down to 2~K. High-pressure studies were performed using the device-integrated diamond anvil cell (DIDAC) technique, featuring pre-patterned electrodes on the diamond anvil~\cite{Xie2021,Ku2022,Zhang2023}. Glycerin served as the pressure-transmitting medium. The pressure was calibrated at room temperature via the ruby fluorescence method.

\begin{figure}[!t]\centering
      \resizebox{8.5cm}{!}{
              \includegraphics{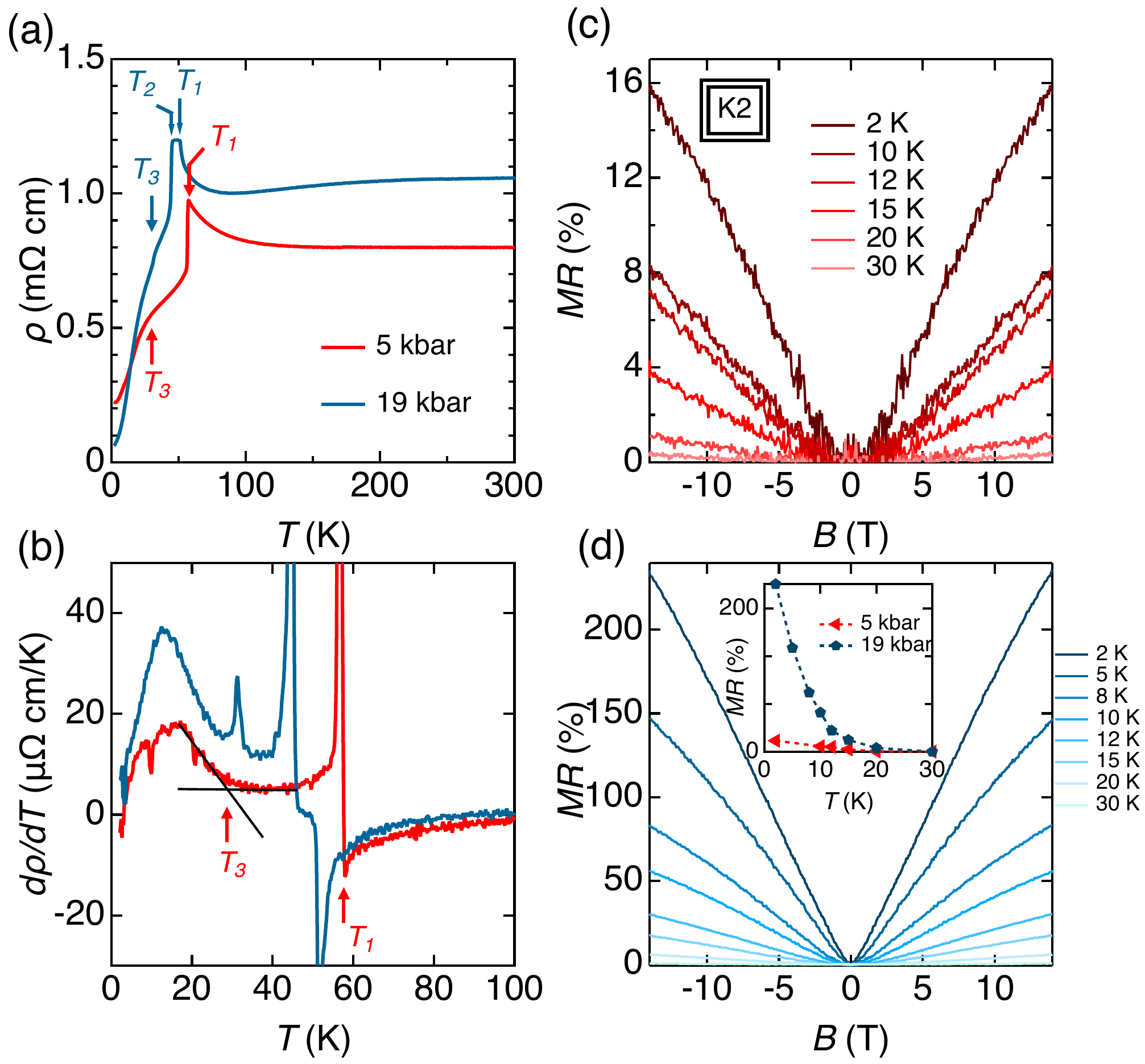}}                				
              \caption{\label{fig1}  
(a) Temperature dependence of the resistivity of \CCS\ (K1) measured at 5~kbar and 19~kbar. Three anomalies are indicated by arrows, labeled as $T_1$, $T_2$, and $T_3$. (b)  Temperature dependence of the derivative of resistivity at 5~kbar and 19~kbar. Two-line intersection method is adopted. $T_1$ and $T_3$ at 5~kbar are indicated by arrows. Magnetoresistance (MR) with \textit{B}~$\parallelsum$ \textit{c} measured at different temperatures at (c) 5~kbar for sample K2 and (d) 19~kbar for sample K1. The inset shows the temperature dependence of MR (14~T) at both pressures. } 
\end{figure}

 %%%%%%result and discussion%%%%%%%%%%%
%%%%%%%FIG 1%%%%%%%%%%%
\indent We first present the temperature dependence of the in-plane resistivity $\rho(T)$ and magnetoresistance (MR=$\frac{\rho(B)-\rho(0)}{\rho(0)}$). As shown in Fig.~\ref{fig1}(a), at near-ambient pressure (5~kbar), $\rho(T)$ remains nearly temperature-independent above 100~K, followed by a pronounced peak at $T_1$ $\approx$ 58~K that corresponds to the previously reported phase transition~\cite{Liu2024}. An anomalous hump appears in $\rho(T)$ accompanied by an inflection point in $d\rho/dT$ as shown in Fig.~\ref{fig1}(b). Since such an anomaly lacks a sharp feature, we used an empirical two-line intersection method to approximate its value and designated it as $T_3$ $\approx$ 30~K with an error bar less than 3~K, as demonstrated in Fig.~\ref{fig1}(b) (see Supplemental Material~\cite{supp} for a detailed discussion). With further cooling, the resistivity decreases steeply with temperature, reaching the residual value $\rho_0$=0.2~m$\Omega \cdot $cm. The residual resistivity ratio (RRR), defined as $\rho(300~\text{K})/\rho(2~\text{K})$, is 3.6. We then apply a magnetic field with $B\parallelsum c$ to study the magnetotransport properties. Figure~\ref{fig1}(c) shows the MR up to 14~T at various temperatures for K2. We find that MR increases significantly at low temperature, reaching 16$\%$ at 2~K. Moreover, MR exhibits sublinear behavior, consistent with that observed in the sister compound \CsVSb~\cite{Yu2021b,zhou2026}. The MR value and behavior of K1 at 5~kbar are almost identical with K2 except a lower signal-to-noise ratio (see Supplemental Material~\cite{supp}).

\indent To trace the transport signatures, we further increase the pressure to 19~kbar. As plotted in Fig.~\ref{fig1}(a), in the high-temperature region ($>$100~K), $\rho(T)$ becomes more metallic compared with the near-ambient curve and exhibits a slight downward tilt. $T_1$ is shifted down to $\approx$ 50~K, accompanied by the emergence of a new feature $T_2$ $\approx$ 46~K, which again agree with the previous reports~\cite{Liu2024}. In contrast, the hump remains observable at around the same temperature $T_3$ $\approx$ 30~K, but its effect on magnetotransport appears to be more significant. Below $T_3$, the resistivity drops even more dramatically, reaching $  \rho_0$=0.06~m$\Omega \cdot $cm, only 30$\%$ of the 5~kbar value. Accordingly, the MR enhancement at low temperature is substantially more pronounced: MR(14~T) reaches 230\% at 2~K, which is 14-fold larger than the near-ambient value at 2~K as shown in the inset of Fig.~\ref{fig1}(d). These observations capture the pressure dependence of $T_3$ and highlight the exotic transport signatures in \CCS\ at low temperatures.

%%%%%%%%%%%%Fig 2%%%%%%%%%%%%%
\begin{figure*}[!t]\centering
      \resizebox{15cm}{!}{
 \includegraphics{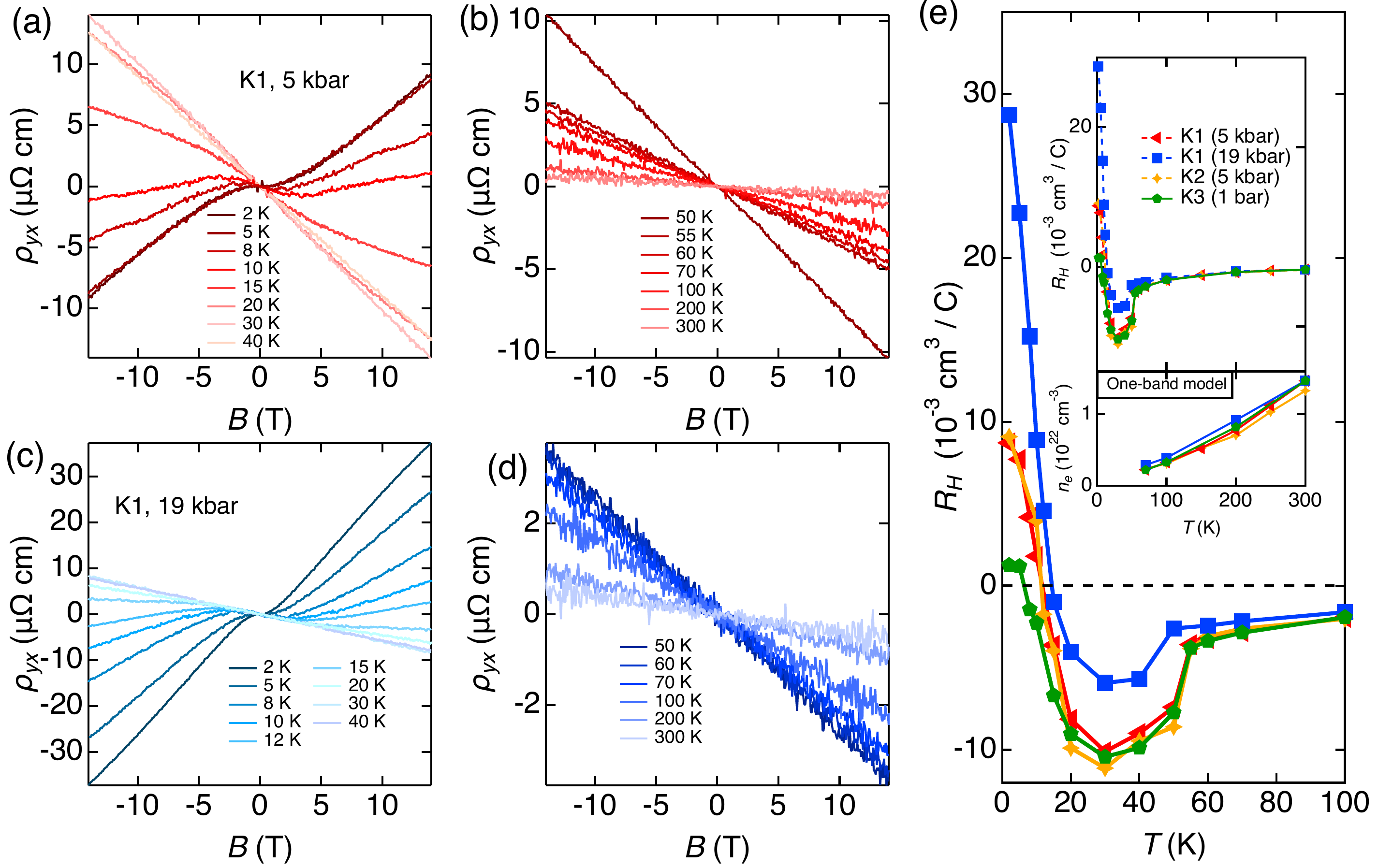}}       
 \caption{\label{fig2} Hall resistivity $\rho_{yx}$ against magnetic field with B $\parallelsum$ c at different temperatures at (a)(b) 5~kbar and (c)(d) 19~kbar for K1. (e) Temperature dependence of $R_H$ for K1 at 5~kbar and 19~kbar, K2 at 5~kbar, and K3 at ambient pressure. The upper inset shows the $R_H$ up to 300~K. The lower inset shows the carrier densities at all pressures extracted using one-band model at high temperatures.}
 
\end{figure*}

To gain deeper insight into the characteristic temperatures detected in the resistivity, we investigated the Hall effect at various temperatures and different pressures. Figures~\ref{fig2}(a) and (b) present the Hall resistivity $\rho_{yx}$ measured from 2~K to 300~K up to 14~T at 5~kbar. At high temperatures, $\rho_{yx}$ exhibits a linear dependence on magnetic field, and the Hall coefficient ($R_H$) becomes progressively more negative upon cooling. Using a simple one-band model, we extract the effective carrier density, as shown in the lower inset of Fig.~\ref{fig2}(e). The substantial drop of the effective carrier density on cooling is surprising, hinting at the needs of considering the multiband effect. Around 55~K, $R_H$ undergoes a sharp drop that continues down to approximately 30~K, as summarized in Fig.~\ref{fig2}(e). This behavior has been previously attributed to the phase transition at $T_1$~\cite{Liu2024}. Below 30~K, $\rho_{yx}$ becomes noticeably non-linear in the low-field regime ($\left|B\right|\leq5$~T), displaying features reminiscent of an anomalous Hall effect. Consequently, we define $R_H$ as the slope of $\rho_{yx}(B)$ extracted from the high-field linear region. Below 30~K, $R_H(T)$ begins to trend toward positive values, exhibiting a pronounced upturn coinciding with the emergence of the anomaly $T_3$ in $\rho(T)$. With further cooling, $R_H(T)$ crosses zero at approximately 15~K and eventually becomes positive, indicating a transition from electron-dominated to hole-dominated charge transport. These intriguing magnetotransport signatures are all reproduced in the other sample K2 (see Supplemental Material~\cite{supp}). This unconventional temperature dependence of $R_H$ suggests complicated carrier characteristics below $T_3$. To examine if this Hall anomaly associated with $T_3$ is pressure-induced, we further measure a separate sample K3 at ambient pressure (see Supplemental Material~\cite{supp}). K3 exhibits the same $R_H(T)$ upturn and sign change below $T_3$ (Fig.~\ref{fig2}(e)), demonstrating that $T_3$ and its associated Hall anomaly are intrinsic properties of \CCS\ even at ambient pressure.

 Similar behavior is observed at 19~kbar, as shown in Figs.~\ref{fig2}(c) and (d), and indicated by the square markers in Fig.~\ref{fig2}(e). At high temperatures, the effective carrier density exhibits an identical temperature dependence but slightly higher value than that in the 5~kbar case, indicating negligible changes in the electronic structure in this temperature region. A sharp drop in $R_H$ emerges at around 50~K, coincident with the suppressed values of $T_1$ inferred from $\rho(T)$ at 19~kbar. This agreement further confirms that the drop is intrinsically linked to the phase transition at $T_1$.
In contrast, the dramatic positive upturn in $R_H(T)$ remains near 30~K at both pressures, coinciding precisely with $T_3$ in $\rho(T)$, which also shows no significant pressure dependence. The close correspondence between these features across different pressures strongly suggests an intimate connection between $T_3$ and the abrupt upturn in $R_H(T)$. Moreover, this upturn in $R_H(T)$ becomes significantly more pronounced at 19~kbar, and the low-field non-linearity appears to be more abrupt, which may reflect an increased influence of $T_3$.
Pronounced turning in the Hall coefficient has been reported in several correlated-electron systems with distinct microscopic origins. For instance, a high-mobility carrier pocket in the nematic phase of FeSe can result in a downturn in $R_H(T)$~\cite{farrar2022, watson2015}; BaFe$_2$As$_2$ shows a drastic upturn in $R_H(T)$ when it enters the spin-density-wave state~\cite{fang2009, iida2016}; YBa$_2$Cu$_3$O$_y$ has a downturn in $R_H(T)$ in a field-enhanced CDW state~\cite{leboeuf2011, badoux2016}; the sister kagome compound \CsVSb\ has a similar upturn in $R_H(T)$ in the CDW state~\cite{Yu2021b}. These examples establish that a pronounced $R_H(T)$ anomaly can indicate a substantial modification to the underlying carrier profile, arising from a variety of distinct microscopic origins. Therefore, these works motivate our identification of $T_3$ as a genuine and notable transport anomaly in \CCS, which merits further investigation using complementary probes.
%%%%%%%%%%%%%Fig 3%%%%%%%%%%%%%%%%
\begin{figure*}[!t]\centering
       \resizebox{15cm}{!}{
     \includegraphics{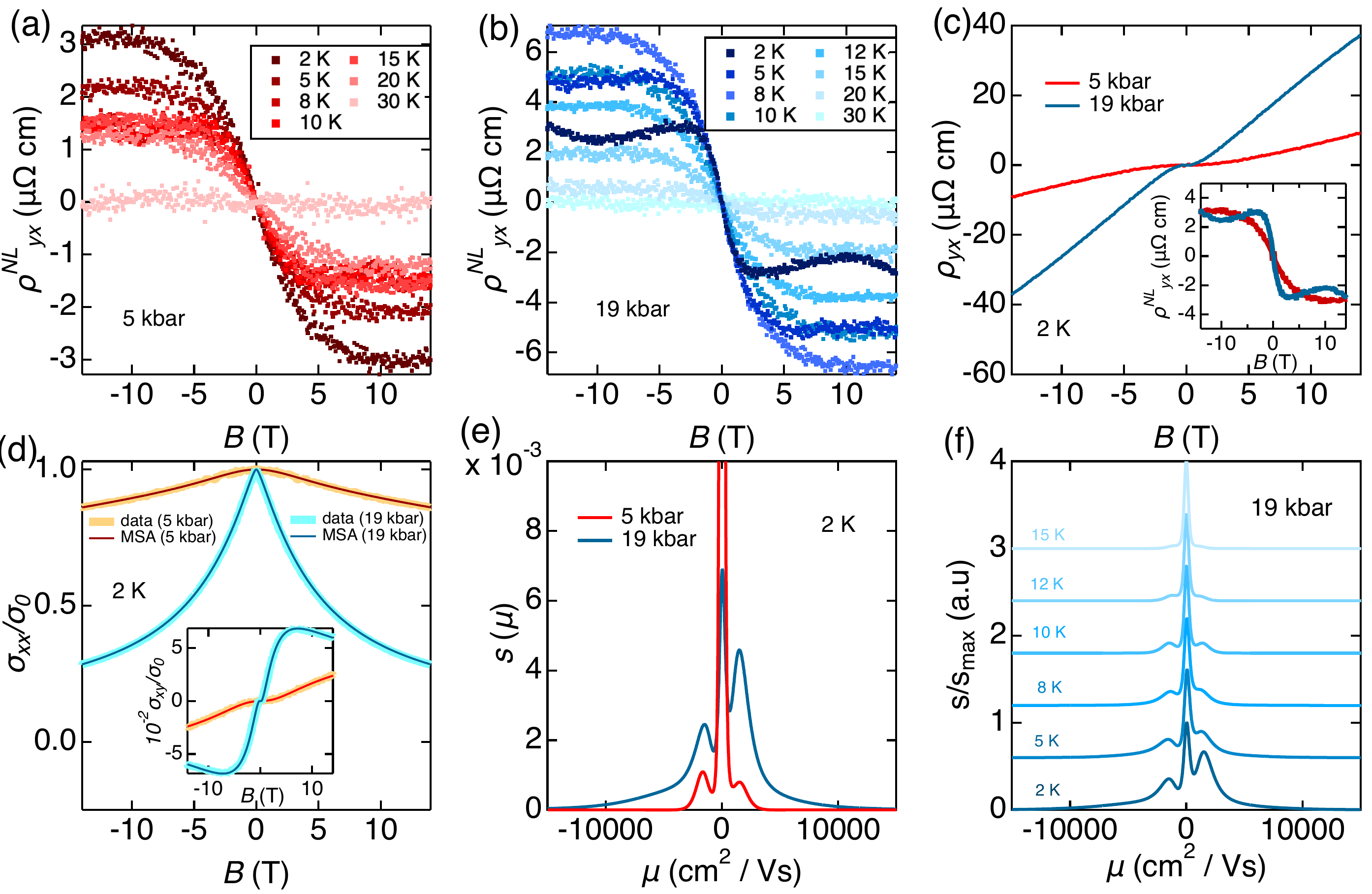}}         
              \caption{\label{fig3} Non-linear component of Hall resistivity $\rho^{NL}_{yx}$ against magnetic field with B $\parallelsum$ c at different temperatures at (a) 5~kbar and (b) 19~kbar. (c) $\rho_{yx}$(B) at 2~K for both pressures. The inset compares the non-linear component at different pressures. (d) Normalized magnetoconductivity and Hall conductivity at 2~K for both pressures, the narrow lines represent the fitting results from Mobility spectrum analysis (MSA). MSA result at (e) 2~K for both pressures and (f) various temperatures at 19~kbar. The spectra are vertically shifted for clarity. }
\end{figure*}

To gain further insights into the carrier characteristics below $T_3$ in \CCS, we study the temperature evolution of AHLE. Following the practice established in the analysis of \AVSb~\cite{Yu2021b,Wang2023,Yang2020}, we isolate the non-linear component by subtracting the linear term extrapolated from the high field region ($\left|B\right|\geq10$~T). As shown in Figs.~\ref{fig3}(a) and \ref{fig3}(b), the non-linear component in $\rho_{yx}(B)$ only emerges below 30~K which again coincides with $T_3$. Such a non-linear Hall response is already present exclusively below $T_3$ at ambient pressure, confirming the robustness of this feature related to $T_3$ (see Supplemental Material~\cite{supp}). More strikingly, $\rho_{yx}(B)$ is more anomalous-Hall-like at 19~kbar compared with 5~kbar case: as exemplified by the comparison of low- and high-pressure datasets at 2~K in Fig.~\ref{fig3}(c), the low-field nonlinearity appears more abrupt with increasing pressure. In addition, the nonlinearity even shows up in the high field region at 19~kbar (see the inset of Fig.~\ref{fig3}(c)). A similar contrast can be found when Hall conductivity is plotted instead, as shown in Fig.~\ref{fig3}(d), where an abrupt change can be discerned at 19~kbar at near-zero field. Such a sharp near-zero-field feature is commonly observed in \AVSb, which has originally been associated with anomalous-Hall effect. Recent analysis, however, re-interpreted the near-zero-field feature in \CsVSb\ as due to high-mobility carriers from small Fermi pockets~\cite{liu2025a}. On the other hand, the nonlinear feature at 5~kbar appears to be rather conventional at near-zero field. Such behavior highlights a distinct change in carrier properties when moving from 5~kbar to 19~kbar.

The mobility spectrum analysis (MSA)~\cite{Tsui2025,Beck1987,Beck2021,Antoszewski1995,Vurgaftman1998,Kiatgamolchai2002,Rothman2006,Wang2025a}, which has played a key role in the re-interpretation of the nonlinear Hall response at near-zero field in \CsVSb~\cite{liu2025a}, is applied to our data for K1 at 5~kbar and 19~kbar. Without the need of assuming the number or types of carriers, the longitudinal and transverse conductivities are given by
\begin{align*}
\sigma_{\rm{xx}}&=\sum_{i}\frac{s(\mu_i)}{1+(\mu_iB)^{2}}~\rm{and}\\
\sigma_{\rm{xy}}&=\sum_{i}\frac{s(\mu_i)\mu_iB}{1+(\mu_iB)^{2}},
\end{align*}
where $i$ represents the discretization steps used for numerical integration, and $s(\mu_i)$ is the partial conductivity of the carrier with mobility $\mu_i$. The detailed fitting procedure can be found in Ref.~\cite{Tsui2025}.

As exemplified by the analysis of the datasets at 2~K (see Fig.~\ref{fig3}(d)), the mobility spectrum at both pressures are presented in Fig.~\ref{fig3}(e) and Supplemental Material~\cite{supp}. The spectra clearly demonstrate the existence of multiple carriers at both pressures, responsible for the observed non-linear Hall response and directly supporting the multi-band characteristic. The peak near zero mobility suggests the presence of unresolved low-mobility carriers that fall below the resolution limit of MSA, see Supplemental Material for discussion~\cite{supp}. However, compared with the 5~kbar data, the MSA result for 19~kbar has a substantially broader mobility spectrum and has a larger contribution from high mobility carriers ($>$ 5,000~cm$^2$/Vs) relative to the low-mobility ones. Focusing on the temperature dependence of the spectrum, it is evident that these high mobility carriers fade out upon warming, as shown in Fig.~\ref{fig3}(f) and Supplemental Material~\cite{supp}. Concomitantly, the AHLE is also weakened upon warming, suggesting a direct link between AHLE and the high mobility carrier~\cite{liu2025a}. 
Thus, the present observation is reminiscent of the case in \CsVSb, in which the possible role of high-mobility carriers on AHLE has been pointed out through MSA. Finally, the close resemblance also hints at the existence of small Fermi pockets below $T_3$ and their active role in magnetotransport. 

While our results provide significant insights into the AHLE in \CCS, other potential origins cannot be definitively ruled out. For example, theoretical studies suggest the presence of an altermagnetic order in \CCS~\cite{Xu2025}, which could generate an internal magnetic field and break time-reversal symmetry (TRS), leading to the AHLE. Another plausible mechanism is the emergence of loop currents inherent to the kagome lattice, a phenomenon long proposed for the \AVSb\ family to explain AHLE~\cite{christensen2022,Zhan2026,tazai2023}. Furthermore, significant Berry curvature may arise directly from the electronic band structure of \CCS\ and break TRS, particularly if the material hosts topologically non-trivial band structures~\cite{Fu2021,Chapai2023}. It is possible that these effects coexist with high-mobility carriers, collectively influencing the observed AHLE.  While one cannot pinpoint a particular origin for AHLE and resolve the mechanism responsible for $T_3$ based solely on magnetotransport measurements, this work establishes a critical foundation that will guide future experimental and theoretical studies toward fully resolving this transport anomaly.

In conclusion, we have conducted systematic investigations on previously overlooked $T_3$ anomaly in \CCS\ via magnetotransport under hydrostatic pressure. Our results provide suggestive evidence for a link between $T_3$ and the emergence of unconventional transport behavior in \CCS. The appearance of enhanced conductivity, the sharp upturn in $R_H(T)$, and the onset of AHLE may point to the activation of additional carrier channels governed by $T_3$. At 19~kbar, the mobility of carriers activated below $T_3$ is further enhanced, leading to markedly reduced residual resistivity and more pronounced anomalous-Hall-like features. While the precise microscopic origin of the anomaly associated with $T_3$ remains to be clarified, its close correspondence with these transport signatures suggests that it plays an important role in the electronic properties of \CCS. Further studies with complementary probes will be valuable in elucidating the nature of these features.

\begin{acknowledgments}
The work at CUHK was supported by the Research Grants Council of Hong Kong (CUHK 14300722, CUHK 14301020,
CUHK 14301725 and CUHK 14302724), CUHK Direct Grant (4053577, 4053664), the Guangdong Provincial Quantum Science Strategic Initiative (GDZX2301009), and the 1+1+1 CUHK–CUHK(SZ)–GDSTC Joint Collaboration Fund (2025A0505000079). The work at BIT was supported by the National Key Research and Development Program of China (Grant Nos. 2025YFA1411200, 2022YFA1403400), and the Beijing National Laboratory for Condensed Matter Physics (Grant No. 2023BNLCMPKF007). Z.W. thanks the Analysis and Testing Center at BIT for assistance in facility support. 
\end{acknowledgments}

%\bibliography{Cr135}

\providecommand{\noopsort}[1]{}\providecommand{\singleletter}[1]{#1}%

\end{document}